\documentclass[a4paper,10pt]{article}

\usepackage{natbib}
\usepackage{graphicx}
\usepackage{subfig}

\makeatletter
\edef\JFMt@xtsize{10}

\renewcommand\normalsize{%
	\@setfontsize\normalsize{\JFMt@xtsize}\@xiipt
	\abovedisplayskip 6.5\p@ \@plus 1\p@ \@minus 1\p@
	\belowdisplayskip \abovedisplayskip
	\abovedisplayshortskip 3\p@ \@plus 1\p@
	\belowdisplayshortskip \abovedisplayshortskip
}

\newlength\halflineskip
\newlength\affilskip
\@beginparpenalty -\@lowpenalty
\@endparpenalty -\@lowpenalty
\@itempenalty -\@lowpenalty
\clubpenalty\z@
\widowpenalty\@M

\makeatother

\usepackage{epstopdf}
\usepackage{caption}

\usepackage{amsmath,amssymb}
\usepackage{mathrsfs}
\usepackage{color}

\newcommand{\Pd}[2][]{\displaystyle \frac{\partial #1}{\partial #2} }
\newcommand{\Pdt}[2][]{\textstyle \partial #1/\partial #2 }

\newcommand{\Od}[2][]{\displaystyle \frac{{\rm d} #1}{{\rm d} #2} }
\newcommand{\Odt}[2][]{\textstyle {\rm d} #1/{\rm d} #2 }

\newcommand{\f}[1]{\overline{#1}}
\newcommand{\ft}[1]{\widetilde{#1}}

\renewcommand{\vec}[1]{\mbox{\boldmath $#1$}}

\begin{document}
	
\title{\bf On the decay of dispersive motions in the outer region of rough-wall boundary layers}
	
	\author{{\bf Johan Meyers$^{1,}$\thanks{Email address for correspondence: johan.meyers@kuleuven.be},~ Bharathram~Ganapathisubramani$^2$~~and}\\ {\bf Ra\'ul~Bayo\'an~Cal$^3$}\\[1em]
		{\normalsize $^1$KU Leuven, Mechanical Engineering, Celestijnenlaan 300, B3001 Leuven, Belgium}\\[0.5em]
		{\normalsize$^2$University of Southampton, Aerodynamics \& Flight Mechanics Group,}\\ {\normalsize Southampton~SO17~1BJ, United Kingdom}\\[0.5em]
		{\normalsize $^3$Portland State University, Mechanical and Materials Engineering,} \\ {\normalsize Portland, OR~97207, United States}}
	
	\date{\today}
	
	\maketitle

\renewcommand{\abstractname}{}
\begin{abstract}
In rough-wall boundary layers, wall-parallel non-homogeneous mean-flow solutions exist that lead to so-called dispersive velocity components and dispersive stresses. They play a significant role in the mean-flow momentum balance near the wall, but typically disappear in the outer layer. A theoretical framework is presented to study the decay of dispersive motions in the outer layer. To this end, the problem is formulated in Fourier space, and a set of governing ordinary differential equations per mode in wavenumber space is derived by linearizing the Reynolds-averaged Navier--Stokes equations around a constant background velocity. With further simplifications, analytically tractable solutions are found consisting of linear combinations of $\exp(-kz)$ and $\exp(-Kz)$, with $z$ the wall distance, $k$ the magnitude of the horizontal wavevector $\vec{k}$, and where $K(\vec{k},\textit{Re})$ is a function of $\vec{k}$ and the Reynolds number $\textit{Re}$. Moreover, for $k\rightarrow \infty$ or $k_1\rightarrow 0$, $K\rightarrow k$ is found, in which case solutions consist of a linear combination of  $\exp(-kz)$ and $z\exp(-kz)$, and are Reynolds number independent. These analytical relations are verified in the limit of $k_1=0$ using the rough boundary layer experiments by Vanderwel and Ganapathisubramani (\textit{J.~Fluid~Mech.}~\textbf{774}, R2, 2015) and are in good agreement for $\ell_k/\delta \leq 0.5$, with $\delta$ the boundary-layer thickness and $\ell_k = 2\pi/k$.

\end{abstract}

\section{Introduction}

When analyzing turbulent flow over rough surfaces, flow statistics are often averaged over wall-parallel planes. The difference between mean flow and horizontally averaged mean flow yields dispersive velocities, and leads to so-called dispersive or coherent stresses in the horizontally averaged mean momentum equations \citep{Raupach1982}. The effects of the shape and distribution of surface roughness on turbulent wall-flows is typically assumed to be confined to the roughness sub-layer where the dispersive motions are dominant and complement Reynolds shear stresses \citep{Raupach1991}. In this region, the mean flow statistics are complex and three dimensional, governed by the complex shape of the roughness itself.

In flows over homogeneous rough surfaces, the roughness sub-layer is typically found to extend vertically 2-5 times the representative roughness heights, which can be the equivalent sandgrain roughness  or the maximum roughness height or a root-mean-square roughness height, depending on the type of roughness \citep{Raupach1991,Nikora2001,Jimenez2004,Flack2007}. When moving away from the wall, the dispersive stress contribution to the total stress gradually decays, in accordance with the picture of an outer layer that is dominated by Reynolds stresses.  Therefore, beyond this roughness sub-layer, the outer layer of the flow is usually independent of local details of surface roughness, resulting in a mean flow that is nearly homogeneous in wall-parallel directions, with flow statistics that mainly depend on the wall-normal direction \citep{Castro2007}. However, for rough surfaces with spatial heterogeneities where dominant spanwise length scales of the roughness distribution are on the order of the outer length scale of the flow, 
large secondary motions are excited by the roughness arrangement, and can penetrate into the outer layer \citep{Nezu1984,Wang2005,Barros2014,Anderson2015,Vanderwel2015,Kevin2017,Medjnoun2018,Hwang2018}. Therefore, dispersive stress can be significant across the entire turbulent layer. 

Recent work has shown that the decay of dispersive stresses (or secondary motions) scales with the spanwise roughness wavelength when the roughness is geometrically scaled (proportional increase in both roughness height and wavelength) and when the wavelength increases at fixed roughness height \citep{Yang2018,Chan2018}. In the current work, a new analytical framework is proposed that allows to study this decay of dispersive motions systematically. The paper is organised as follows. First in Section~\ref{s:dispersivesolutions}, the theory is presented, and approximate solutions for dispersive motions in the outer layer of a boundary layer are derived. Next, a comparison with experiments is shown in Section~\ref{s:experiments}. Lastly, discussion and conclusions are stated in Section~\ref{s:discussion}. 

\section{Approximate solutions for the dispersive velocity field}\label{s:dispersivesolutions}
\subsection{Horizontally-averaged Navier--Stokes equations and linearization}
Consider an incompressible turbulent boundary layer over a rough wall, with $x_1$ oriented in streamwise, $x_2$ in spanwise, and $x_3$ in wall-normal direction. Further, $\f{\vec{u}}$ represents the Reynolds-averaged velocity field, with fluctuation $\vec{u}'$. The focus is on rough boundary layers with either periodic roughness elements or a roughness distribution that is statistically homogeneous in horizontal directions, and the horizontally averaged mean flow is denoted with $\langle \f{\vec{u}} \rangle \equiv (U,0,W)$.  Furthermore, $\f{\vec{u}}''$ is introduced, so that $\f{\vec{u}} = U \vec{e}_1 + W \vec{e}_3 + \f{\vec{u}}''$.  

It is further presumed that the boundary layer is sufficiently developed for the streamwise evolution of mean velocity components to be negligible, so that the time-averaged and horizontally averaged Navier--Stokes equation follows as
\begin{align}
W \Pd[U]{x_3}  - \nu \Pd[^2 U]{x_3^2} + \frac{1}{\rho}\Pd[p_\infty]{x_1} = - \Pd[\langle\f{u_1'u_3'}\rangle]{x_3} - \Pd[\langle\f{u}_1''\f{u}_3''\rangle]{x_3} \label{eq:planeavNS}
\end{align}
with $\langle\f{u_1'u_3'}\rangle$ and $\langle\f{u}_1''\f{u}_3''\rangle$ being the plane averaged Reynolds stress and dispersive stress respectively, $p_\infty$ the background pressure, $\nu$ the kinematic viscosity, and where the density $\rho$ is presumed to be constant. 

The equations for the dispersive velocity fluctuations $\f{\vec{u}}''$ further follow from subtracting (\ref{eq:planeavNS}) from the standard Navier--Stokes equations, yielding
\begin{align}
&\Pd[\f{u}_i'']{x_i} = 0, \label{eq:Continuitydispersive}\\
&U \Pd[\f{u}_i'']{x_1} + \f{u}_j'' \Pd[\f{u}_i'']{x_j} + \f{u}_3'' \Gamma \delta_{i1} = -\frac{1}{\rho}\Pd[\f{p}'']{x_i} + \Pd[\langle\f{u_1'u_3'}\rangle]{x_3} \delta_{i1} + \Pd[\langle\f{u}_1''\f{u}_3''\rangle]{x_3} \delta_{i1} - \Pd[\f{u_i'u_j'}]{x_j} + \nu \Pd[^2 u''_i]{x_jx_j}.\label{eq:NSdispersive}
\end{align}
Here, terms with products of $W$ and $\f{u}''_i$ are considered as being negligible, and used the short-hand notation $\Gamma = \Pdt[U]{x_3}$.

For $\f{\vec{u}}''$ sufficiently small, the dispersive velocity equations can be linearized around the mean background flow, neglecting all higher order terms. This leads to 
\begin{align}
\Pd[\f{u}_i'']{x_i} = 0, \label{eq:linearizedcontinuity} \\
U \Pd[\f{u}_i'']{x_1} + \f{u}_3'' \Gamma \delta_{i1} = -\frac{1}{\rho}\Pd[\f{p}'']{x_i} - \Pd[(\f{u_i'u_j'})'']{x_j} + \nu \Pd[^2 u''_i]{x_jx_j} . \label{eq:linearizedmomentum}
\end{align}
In particular, the solution of these equations are investigated in the outer layer, since we expect $\f{\vec{u}}''$ to be small for $x_3\rightarrow \delta$, with $\delta$ the boundary layer thickness.

\subsection{Turbulence closure}
In order to solve (\ref{eq:linearizedmomentum}), a closure is required for the Reynolds stresses $R''_{ij}= (\f{u_i'u_j'})''$. To this end, first, a simple closure for the background flow is posed. Providing that conditions for the linearization hold, it is reasonable to assume that the background corresponds to a standard outer layer solution in the absence of any dispersive terms.  Pertaining to the Reynolds forces, an exact parametrization then corresponds to
\begin{equation}
F_{i} \triangleq -\Pd{x_j}\left(R_{ij} -\frac13 \delta_{ij} R_{kk} \right)  = \Pd[]{x_j} \nu_e \left(\Pd[\f{u}_i]{x_j} + \Pd[\f{u}_i]{x_j}\right) = \delta_{i1} \Pd[\nu_e  \Gamma]{x_3} ,
\end{equation}
where as usual, the trace of the Reynolds stress is absorbed in the pressure term, and the eddy viscosity $\nu_e$ is straightforwardly determined from $\nu_e = -\langle \f{u_1' u_3'} \rangle/\Gamma$, using known experimental, numerical or analytical profiles for $\Gamma$ and $\f{u_1' u_3'}$ in the outer layer of a boundary layer. For later use in \S\ref{ss:Fourier}, the total viscous force is introduced as $G_i\triangleq F_i + \Pdt{x_j} [\nu (\Pdt[\f{u}_i]{x_j} + \Pdt[\f{u}_i]{x_j})]$.

The linearization of the Reynolds forces now follows from the chain rule as
\begin{equation}
F_{i}'' = \Pd[]{x_j} \left[ \nu_e \left(\Pd[\f{u}''_i]{x_j} + \Pd[\f{u}''_i]{x_j}\right) + \delta_{i1} \delta_{j3} \nu''_e  \Gamma \right]. \label{eq:Reynolds_forces}
\end{equation}
Unfortunately, the dispersive turbulent viscosity $\nu''_e$ is not known. However, when $\Gamma \ll 1$, which is generally true in boundary layers for $x_3 \rightarrow \delta$, the term with $\nu''_e  \Gamma$ disappears. Alternatively, this result is also obtained by linearizing the Navier--Stokes equations around a constant background velocity $U_\infty$ instead of $U$. The resulting equations are the same as (\ref{eq:linearizedmomentum}), but with $U_\infty$ instead of $U$ and $\Gamma = 0$. The linearization in that case is valid, as long as $|U_\infty-U| \ll 1$, which again holds for $x_3 \rightarrow \delta$. Thus, for this particular case, (\ref{eq:Reynolds_forces}) with $\Gamma=0$ yields an exact closure of the linearized Reynolds forces. Finally, remark that the above proposed closure is not exact in terms of the Reynolds stresses themselves. This closure determines the Reynolds stresses up to an addition of a divergence-free tensor. The latter will have no influence on the Reynolds force, but will change the individual stress components. In particular, it is well understood that a classical eddy viscosity model leads to a stress tensor with a zero diagonal (corresponding to $\f{u_1' u_1'}=\f{u_2' u_2'}=\f{u_3' u_3'}$), which is generally not the case in boundary layers. Here, this is not an issue, as the Reynolds stresses are not directly needed in the remainder of this work.

\subsection{Representation using Fourier modes}\label{ss:Fourier}
Given periodic roughness elements, the linearized Navier--Stokes equations can be solved using periodic boundaries in $x_1$ and $x_2$ directions. Thus, solutions can be expressed based on a Fourier series. To this end,
\begin{align}
\f{u}_i'' = \sum_{\vec{k}} \ft{u}_i(\vec{k},z) \ \exp(\imath (k_1 x_1 + k_2 x_2)), \label{eq:Fourierspace1}\\
\f{p}'' = \sum_{\vec{k}} \ft{p}(\vec{k},z) \ \exp(\imath (k_1 x_1 + k_2 x_2)), \label{eq:Fourierspace2} 
\end{align}
etc., are introduced with  $\vec{k}=(k_1,k_2)$, and $z\triangleq x_3$. Furthermore, $k_1 = i 2\pi/L_x$, $k_2=j 2\pi/L_y$, with $i,j\in \mathbb{Z}$. Remark that in the case of a roughness distribution, which is statistically homogeneous in horizontal directions (instead of periodic roughness elements), the above Fourier series can be replaced by Fourier integrals in horizontal planes, without further affecting results below. In this case, it is assumed that the largest horizontal length scales in the roughness distributions are sufficiently small for the streamwise homogeneity assumption of the dispersive-flow equations (\ref{eq:Continuitydispersive}, \ref{eq:NSdispersive}) to hold.

Since solving linear equations is the aim, solutions can now be found mode by mode. To this end, the continuity equation is first eliminated by using $\ft{u}_3$ and $\ft{\omega}_3 =  - \imath k_2 \ft{u}_1 + \imath k_1 \ft{u}_2 $ as independent variables. Thus for $(k_1,k_2) \neq (0,0)$,
\begin{align}
\ft{u}_1 &= \frac{\imath k_1}{k^2} \Od[\ft{u}_3]{z} + \frac{\imath k_2}{k^2} \ft{\omega}_3, \label{eq:expressu1} \\
\ft{u}_2 &= \frac{\imath k_2}{k^2} \Od[\ft{u}_3]{z} - \frac{\imath k_1}{k^2} \ft{\omega}_3, \label{eq:expressu2}
\end{align}
with $k = (k_1^2+k_2^2)^{1/2}$. For $k_1=k_2=0$, $\ft{u}_2=\ft{u}_3=0$ is simply found, while $\ft{u}_1 = 0$ in case of linearization around $U$ or $U-U_\infty$ in case of linearization around $U_\infty$. 

Inserting (\ref{eq:expressu1}) and (\ref{eq:expressu2}) in the linearized momentum equations (\ref{eq:linearizedmomentum}), and further eliminating the pressure, then leads to following set of equations (see Appendix A for details)
\begin{align}
- k_1 U \ft{\omega}_3 + k_2 \ft{u}_3 \Gamma =&  k_2 \tilde{G}_1 - k_1 \tilde{G}_2, \label{eq:coupledeq1}  \\
k_1  U \Od[^2\ft{u}_3]{z^2} -  k_1  \ft{u}_3 \Od[\Gamma]{z} - k_1  U k^2  \ft{u}_3  =&  -k_1 \Od[\tilde{G}_1]{z} -  k_2 \Od[\tilde{G}_2]{z} + \imath k^2 \tilde{G}_3, \label{eq:coupledeq2}
\end{align}
which constitutes a set of two coupled ordinary differential equations. Finally, using (\ref{eq:Reynolds_forces}) and $\Gamma=0$  into above equations, using  $U\approx U_\infty$, and some straightforward but cumbersome algebraic manipulations (see Appendix A), leads to
\begin{align}
&\Od{z} \nu_t \Od[\ft{\omega}_3]{z}  -  (\imath U_\infty k_1 + \nu_t k^2) \ft{\omega}_3 = 0 \label{eq:finaleqomega} \\
&\Od[^2]{z^2} \nu_t \Od[^2\ft{u}_3]{z^2} -  \Od[]{z}\left[ (\imath k_1  U_\infty + 2 k^2\nu_t) \Od[\ft{u}_3]{z}\right] +  \left(\imath k_1  U_\infty  + k^2 \nu_t + \Od[^2 \nu_t]{z^2} \right) k^2 \ft{u}_3 = 0 \label{eq:finalequ3},
\end{align}
with $\nu_t = \nu + \nu_e$ the total viscosity. For a classical developing boundary layer, boundary conditions at $z=\infty$ correspond to $\ft{\omega}_3(\infty)=0$, $\ft{u}_3(\infty)=0$, and $\Odt[\ft{u}_3]{z}|_{z=\infty} = 0$. Three more boundary conditions are required to uniquely determine solutions. They should be given at a location $z$ which is sufficiently far from the wall for the linearized equations to hold. These additional conditions are not \textit{a priori} known, and depend on the shape of the wall roughness, and the nonlinear dynamics of the flow close to the wall. 

\subsection{Analytical solutions}\label{s:analytical}
Given an outer layer parametrization of $\nu_e(z)$, and appropriate boundary conditions,  (\ref{eq:finaleqomega}, \ref{eq:finalequ3}) can be solved. Here, the approach is however further simplified by considering a constant eddy viscosity, for which solutions are analytically tractable. 

When  $\nu_t=\nu+\nu_e$ is constant, (\ref{eq:finaleqomega}, \ref{eq:finalequ3}) simplify to
\begin{align}
& \Od[^2\ft{\omega}_3]{z^2}  -  (\imath U_\infty k_1/\nu_t + k^2) \ft{\omega}_3 = 0, \\
& \Od[^4\ft{u}_3]{z^4} - (\imath k_1  U_\infty/\nu_t + 2 k^2)  \Od[^2\ft{u}_3]{z^2} +  \left(\imath k_1  U_\infty/\nu_t  + k^2 \right) k^2 \ft{u}_3 = 0. 
\end{align}
The first equation has two characteristic roots, i.e $\pm K = \pm k (1 + \imath Uk_1/k^2/\nu_t)^{1/2}$, or elaborated in its real and imaginary parts:
\begin{align}
K &= k \left[\sqrt{\frac12\left(1+\left(\frac{U_\infty k_1}{\nu_tk^2}\right)^2\right)^{1/2} + \frac12} ~~ + ~~ \imath \sqrt{\frac12\left(1+\left(\frac{U_\infty k_1}{\nu_tk^2}\right)^2\right)^{1/2}-\frac12} \right]. \label{eq:expressionK}
\end{align}
The second equation has four characteristic roots, i.e. $\pm k$ and $\pm K$. The root $K$ depends on $U_\infty k_1/(\nu_tk^2)$. Introducing $\ell_k \triangleq 2\pi/k$, it can be elaborated as
\begin{equation}
\frac{U_\infty k_1}{\nu_tk^2} = \frac{U_\infty \delta}{\nu_t}  \frac{\ell_k}{2\pi \delta} \frac{k_1}{k} = \left[ \sqrt{\frac{c_f}{2}} \frac{\nu_e}{u_\tau \delta} + \frac{1}{\textit{Re}} \right]^{-1}  \frac{\ell_k}{2\pi \delta} \frac{k_1}{k}, 
\end{equation}
with the Reynolds number $\textit{Re}\triangleq U_\infty \delta/\nu$, and $c_f$ the skin friction coefficient. For $\textit{Re}\rightarrow \infty$, it is expected that $\nu_e/(u_\tau \delta)=\mathcal{O}(1)$, so that $U_\infty k_1/(\nu_tk^2)\sim c_f^{-1/2} (\ell_k/\delta)(k_1/k)$. Other interesting limits correspond to $k_1/k \rightarrow 0$ and $\ell_k/\delta \rightarrow 0$. In both cases, $U_\infty k_1/(\nu_tk^2)\rightarrow 0$, and $K \rightarrow k$ (independent of Reynolds number).

Using the boundary conditions at $z=\infty$, and presuming $U_\infty k_1/k^2\neq 0$, solutions correspond to
\begin{align}
\ft{\omega}_3 &= A \exp(-K z), \quad \mbox{and}~~ \ft{u}_3 = B \exp(-kz) + C \exp(-K z), \label{eq:solutionomw}
\end{align}
with $A$, $B$, $C$ complex numbers that can only be determined if additional boundary conditions are known.
For $U_\infty k_1/k^2 = 0$, $K=k$, and solutions correspond to
\begin{align}
\ft{\omega}_3 &= A \exp(-k z), \quad \mbox{and}~~ \ft{u}_3 = B z \exp(-kz) + C \exp(-k z). \label{eq:solutionk10omw}
\end{align}
Moreover, for the limit of $k_1/k \rightarrow 0$ or $\ell_k/\delta \rightarrow 0$, (\ref{eq:solutionomw}) converges to (\ref{eq:solutionk10omw}).

\section{Experimental verification}\label{s:experiments}

\begin{figure}
	\begin{center}
		\includegraphics[width=\textwidth]{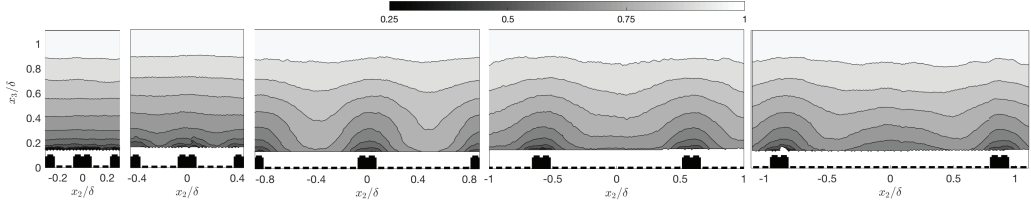}
	\end{center}
	\caption{Overview of the experiments by \cite{Vanderwel2015}. Left to right: contours of streamwise velocity $U/U_\infty$ shown in a spanwise--wall-normal plane obtained from the different experiments. Roughness elements are aligned in streamwise direction and have a different spanwise spacing per experiment. The shape of the elements is outlined at the bottom of the pictures.}\label{f:overviewexp}
\end{figure}

In order to evaluate the relations derived above, experiments by \cite{Vanderwel2015} are examined. These experiments consist of a series of PIV measurements in rough-wall boundary layers with with lego-brick roughness elements that are periodically organized with different spanwise spacings. An overview of the experiment and some mean velocity fields is shown in figure~\ref{f:overviewexp}. The roughness structure is such that $k_1=0$, while spanwise spacings of the different experiments correspond to $S/\delta=0.3$,  $S/\delta=0.45$, $S/\delta=0.88$, $S/\delta=1.2$, $S/\delta=1.8$, with $S$ the wavelength and $\delta$ the boundary layer thickness. Velocity measurements are obtained in a wall-normal--spanwise plane, which allows to fully characterize the dispersive velocity field (given $k_1=0$). Full details of the experiment are found in \cite{Vanderwel2015}.

The PIV measurement planes in the experiments extended approximately 240~mm in spanwise direction. The extent of these planes does not exactly correspond to an integer multiple of the spanwise roughness spacing (i.e, $S=$32~mm, 48~mm, 96~mm, 128~mm, and 192~mm) in the experiments. Therefore, the outer portions of the plane are truncated, and the velocity and Reynolds stress measurements are resampled in spanwise direction so that an integer number of spanwise periods and grid points is retained. Linear interpolation is used for the resampling; keeping the resolution as close as possible to the original. For the different cases ($S/\delta=0.3$,  $0.45$, $0.88$, $1.2$, $1.8$), this yields periods of 7, 4, 2, 1, and 1, respectively. These results are then averaged in spanwise direction and used to obtain the dispersive stress fields as function of wall-normal direction.

First of all, in figure~\ref{f:overview_stresses} an overview is provided of the total stresses as function of wall distance for the different experiments, as well as of the $\langle \f{u}_1'' \f{u}_1'' \rangle$, $\langle \f{u}_3'' \f{u}_3'' \rangle$, and $\langle \f{u}_1'' \f{u}_3'' \rangle$ dispersive stresses. Moreover, based on the maximum of the total stress, the skin friction coefficients for the various cases are also estimated. Figure~\ref{f:overview_stresses} shows that the total stress depends non-monotonously on the spanwise roughness spacing, with the lowest skin friction at $S/\delta=0.3$, which increases to a maximum around  $S/\delta=0.88$, and subsequently decreases again when $S/\delta$ is further increased to $1.8$. Similar differences in skin friction as function of spanwise spacing were, e.g., also observed by \cite{Medjnoun2018, Hwang2018}. 

\begin{figure}[t]
	\begin{center}
		\includegraphics[width=0.42\textwidth]{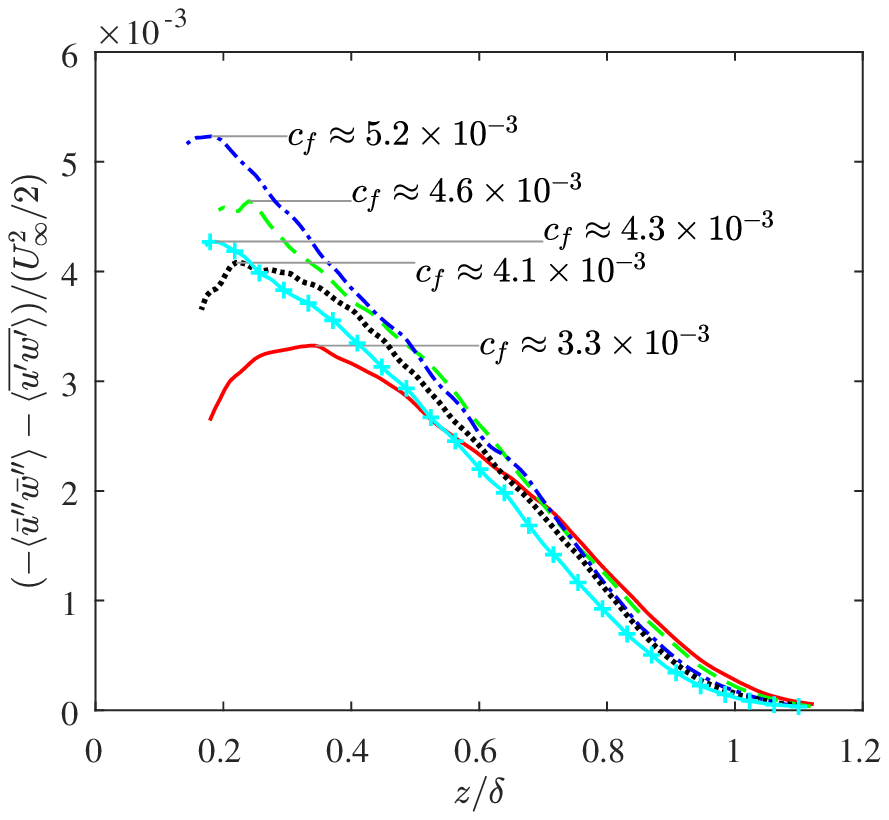}{\small (a)}
		\includegraphics[width=0.42\textwidth]{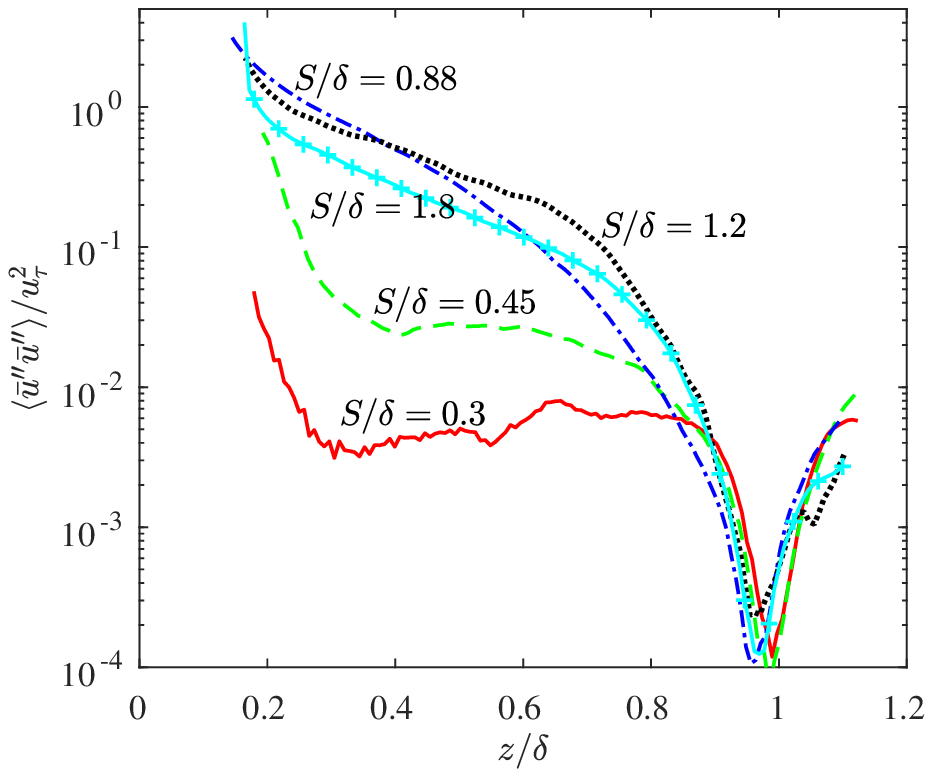}{\small (b)}
		\includegraphics[width=0.42\textwidth]{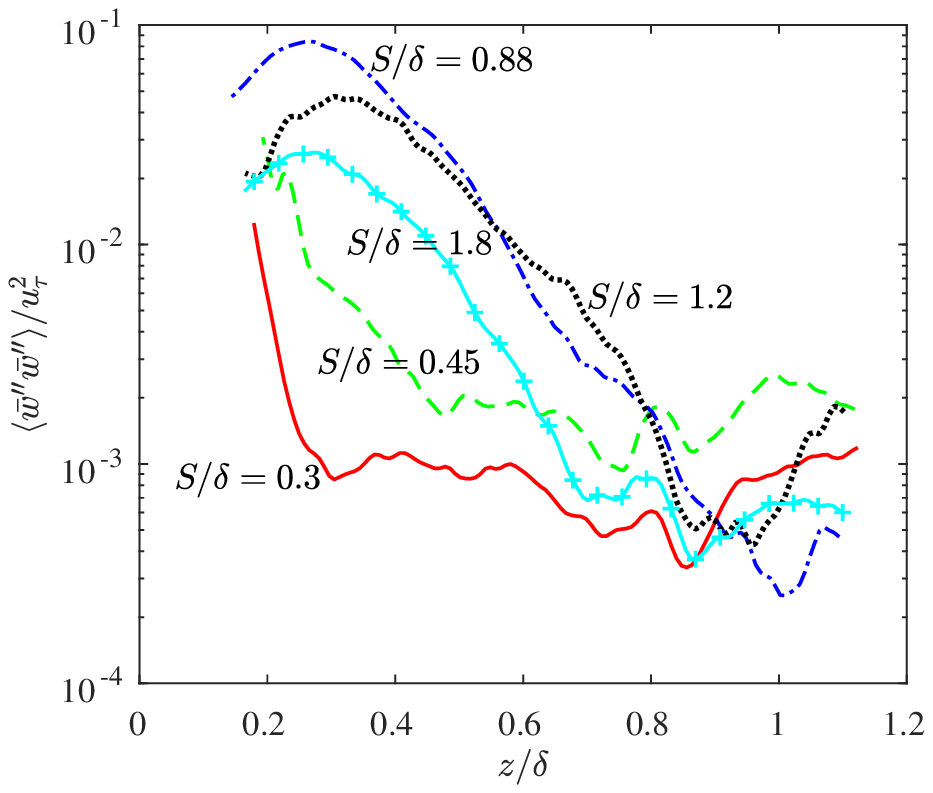}{\small (c)}
		\includegraphics[width=0.42\textwidth]{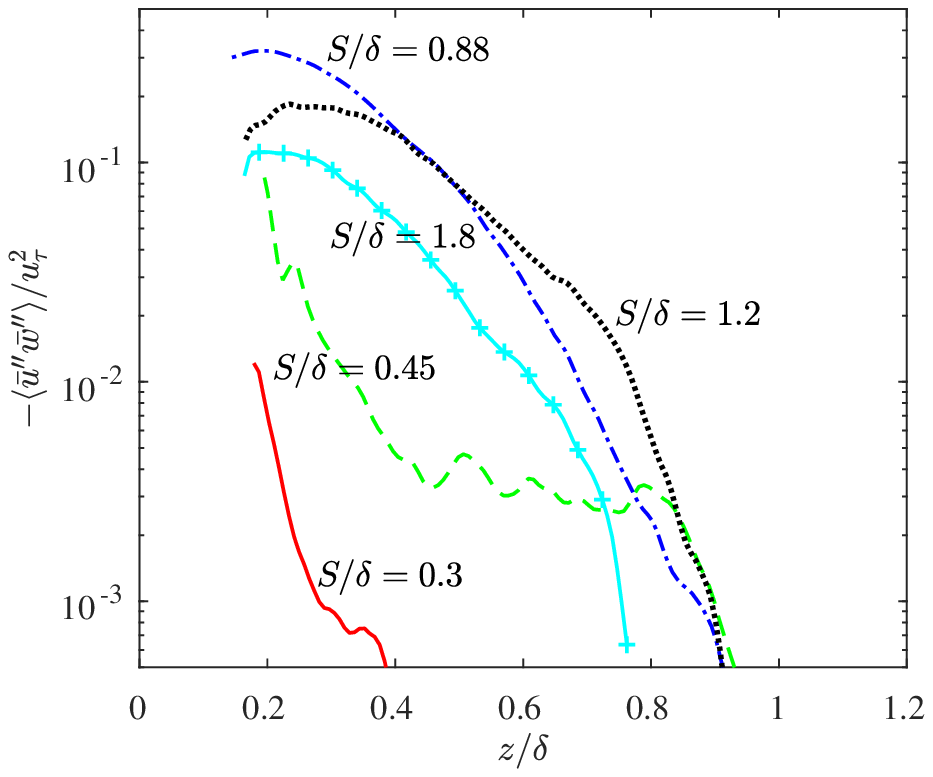}{\small (d)}
	\end{center}
	\caption{(a) Total stress as function of wall distance, normalized by dynamic pressure. (b,c,d) $\langle \f{u}_1'' \f{u}_1'' \rangle$, $\langle \f{u}_3'' \f{u}_3'' \rangle$, and $-\langle \f{u}_1'' \f{u}_3'' \rangle$ respectivelly, as function of wall distance. Linetypes ( {\color{red}---},{\color{blue}$--$},{\color{green}$-\cdot$},$\cdots$,{\color{cyan}--$+$}) correspond to different spacings.}\label{f:overview_stresses}
\end{figure}

Further in figure~\ref{f:overview_stresses}(c--d), it is observed that the maximum magnitude of the dispersive stresses (observed around $z/\delta\approx 0.2$) is correlated with the total skin friction. In particular, for the shear stress, the relative importance of the dispersive stress is largest for the case with largest skin friction (i.e. with $S/\delta=0.88$). These differences result from processes induced by the shape and spacing of the roughness elements as well as the flow in the inner layer of the boundary layer. They essentially serve as a boundary condition for the outer layer decay relations derived in \S\ref{s:dispersivesolutions}.

Finally, it is seen in figure~\ref{f:overview_stresses}(c--d) that the dispersive stresses roughly decay exponentially with increasing wall-normal distance as suggested by the analysis in \S\ref{s:dispersivesolutions}. However, there is no clear single slope to be identified, as the stresses result from a sum over all the modes present in the flow; each decaying at their own rate. To this end, the decay should be analyzed mode by mode (cf. below). It is further acknowledged that the decay saturates towards the top of the boundary layer, as measurement noise starts to play an important role in the very small remaining dispersive stress components. In fact, based on the results in figure~\ref{f:overview_stresses}(c--d), the dynamic range on $\langle \f{u}_i'' \f{u}_j'' \rangle$ is estimated to be on the range of two to three decades. This is in agreement with measurement accuracy of 1\% $U_\infty$ on the velocity field as reported by \cite{Vanderwel2015}. 

Now turning to the evaluation of the decay of dispersive stresses per mode, a Fourier transform in spanwise direction on the dispersive velocity field $\f{\vec{u}}''$ is performed. The decay of the spectra of stream and wall-normal dispersive velocity components is the focal point, respectively defined as $S_{uu}(k_2,z) = \f{u}_1(k_2,z)\f{u}^*_1(k_2,z)$ and $S_{ww}(k_2,z) = \f{u}_3(k_2,z)\f{u}^*_3(k_2,z)$, where $*$ is used for the complex conjugate. Given the relations (\ref{eq:solutionk10omw}) and (\ref{eq:expressu1}), 
\begin{align}
S_{uu} = D_1 \exp(-2 k_2 z), \label{eq:Suuexpect}\\
S_{ww} = \exp(-2 k_2 z) (D_2 + D_3 z + D_4 z^2), \label{eq:Swwexpect}
\end{align}
is expected, where $D_1$--$D_4$ are constants that can, e.g., be found by matching the linear solution to the near wall nonlinear solution at a location sufficiently far from the wall.

\begin{figure}[b!]
	\begin{center}
		\includegraphics[width=0.42\textwidth]{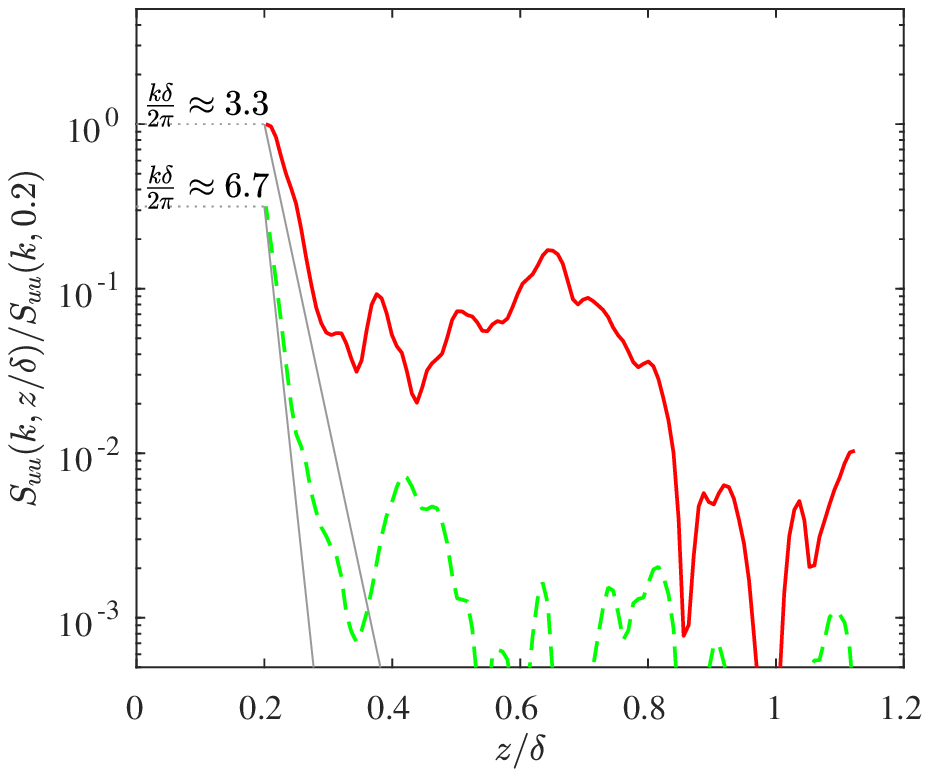}
		\includegraphics[width=0.42\textwidth]{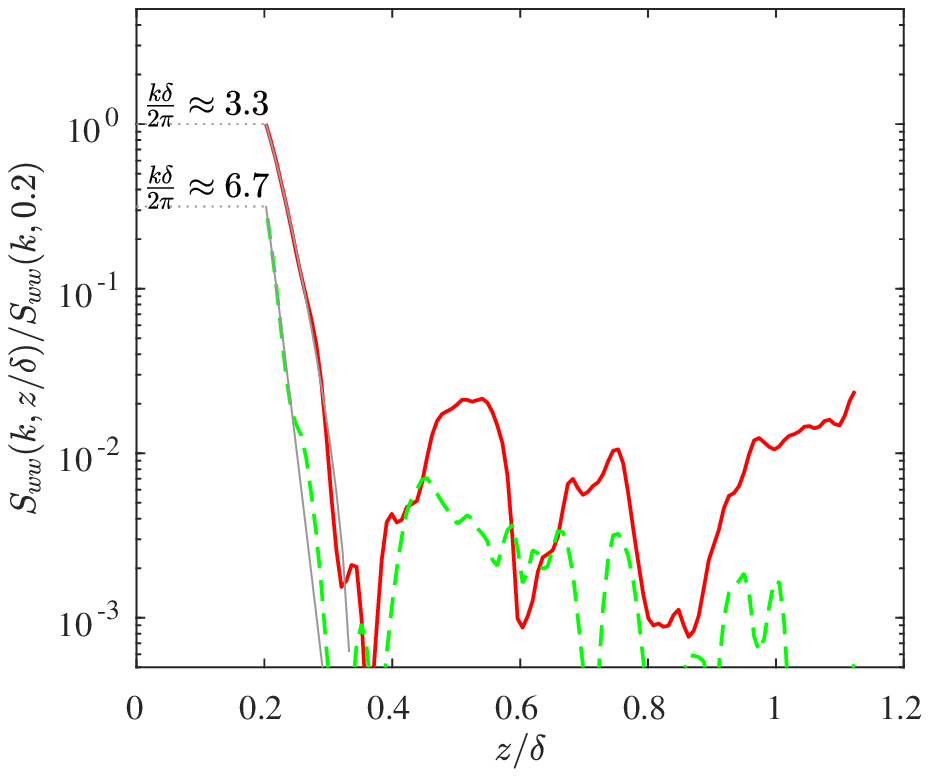}{\small (a)}
		\includegraphics[width=0.42\textwidth]{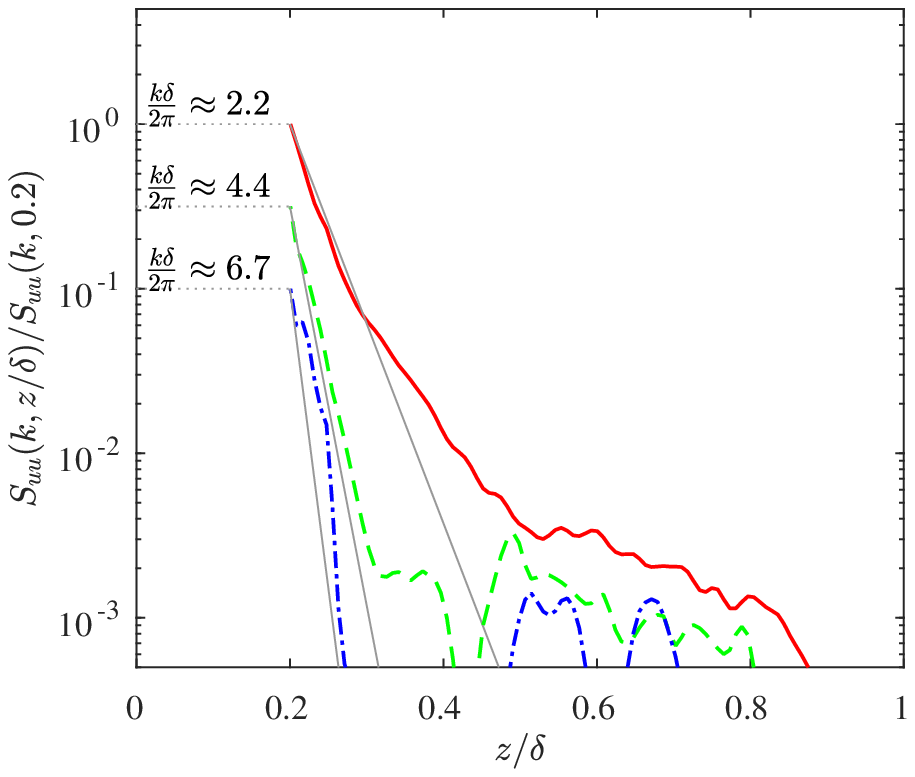}
		\includegraphics[width=0.42\textwidth]{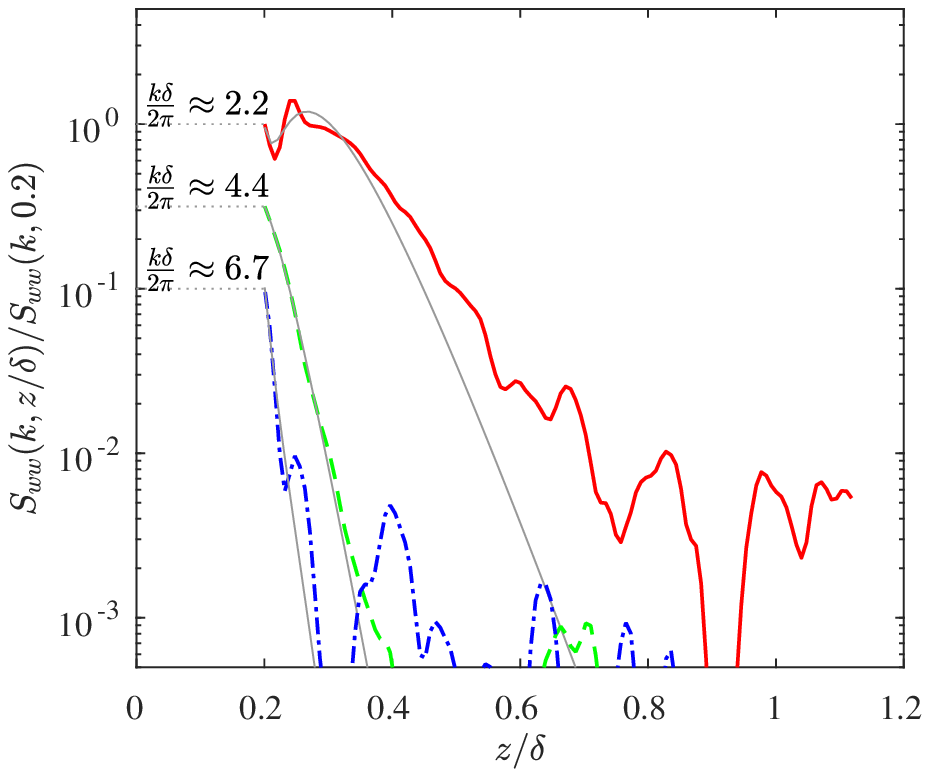}{\small (b)}
		\includegraphics[width=0.42\textwidth]{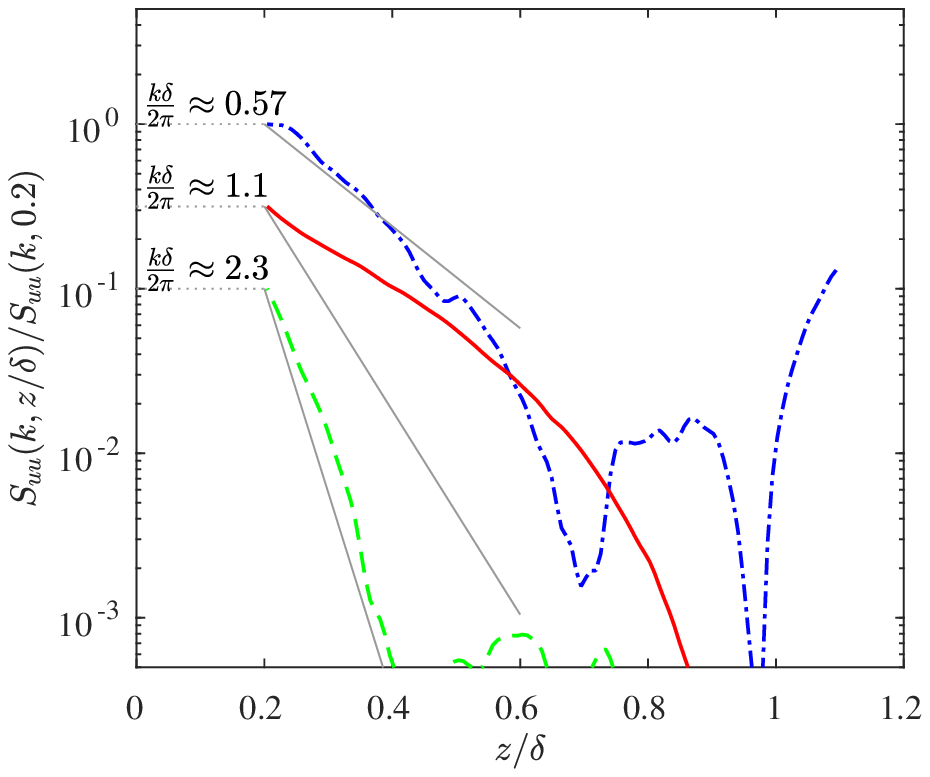}
		\includegraphics[width=0.42\textwidth]{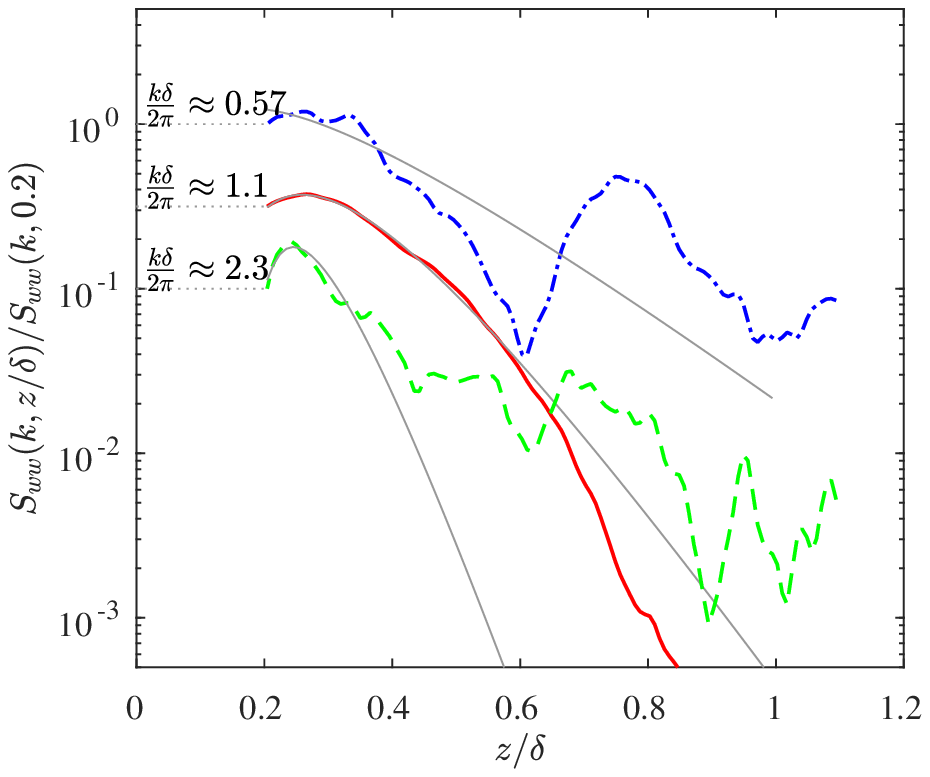}{\small (c)}
	\end{center}
	\caption{(continued on next page) Decay of Fourier modes of streamwise (left) and normal (right) dispersive components as function of wall distance $z/\delta$. (a) $S/\delta=0.3$ (b) $S/\delta=0.45$ ; (c) $S/\delta=0.88$; (d) $S/\delta=1.2$;  (e) $S/\delta=1.8$. ({\color{red}---},{\color{blue}$--$},{\color{green}$-\cdot$},{\color{cyan}$\cdots$}): up to four Fourier modes with highest contribution to the dispersive shear component $\langle f{u}_1'' f{u}_3'' \rangle$; line types are ordered according to magnitude of the contributing mode. Lines are shifted with half a decade for visibility. (gray line in left plots): slope corresponding to $\exp(-2k_2z)$. (gray line in right plots): least-squares fit of $\exp(-2k_2z) (D_2  + D_3 z + D_4 z^2)$ to the data in the range $0.2<z<1.0$.}
\end{figure}

In figure~\ref{f:modedecay}, decay of $S_{uu}$ (left panels) and $S_{ww}$ (right panels) is shown as function of wall distance for the different cases (parts a--e). Results are shown in semi-log scale, and up to four modes are shown, corresponding to those modes that most contribute to the $\langle \f{u}_1'' \f{u}_3'' \rangle$ dispersive stress at $z=0.2\delta$. Next to that, in the left panels the slopes $-2k_2$ (corresponding to (\ref{eq:Suuexpect})) are also plotted for the different modes, while in the right panel least-squares fits of (\ref{eq:Swwexpect}) over the range $0.2<z/\delta < 1.0$ are shown. Remark that in figure~\ref{f:modedecay}(a--c) only two and three modes are shown, as other contributing modes fall within the noise level of the measurements. Similar to before, a dynamic range of two to three decades in the spectra is observed, but for modes with less energy content, the dynamic range can be significantly lower (see, e.g., mode $k\delta/(2\pi)=5.1$ in figure~\ref{f:modedecay}(e)).

\begin{figure}\ContinuedFloat
	\begin{center}
		\includegraphics[width=0.42\textwidth]{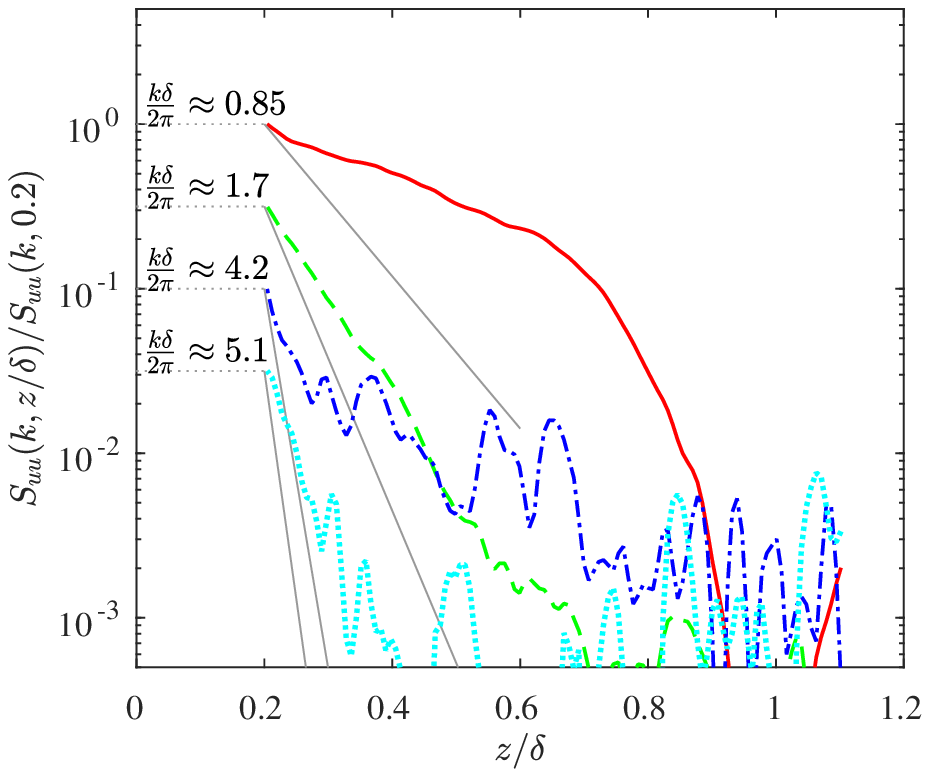}
		\includegraphics[width=0.42\textwidth]{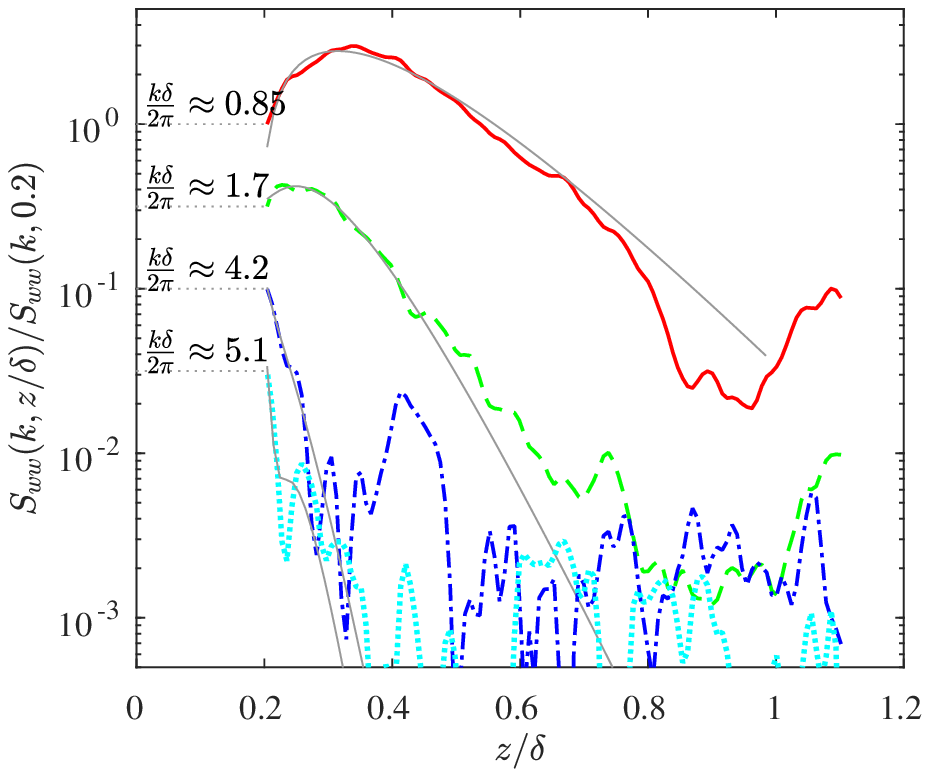}{\small(d)}
		\includegraphics[width=0.42\textwidth]{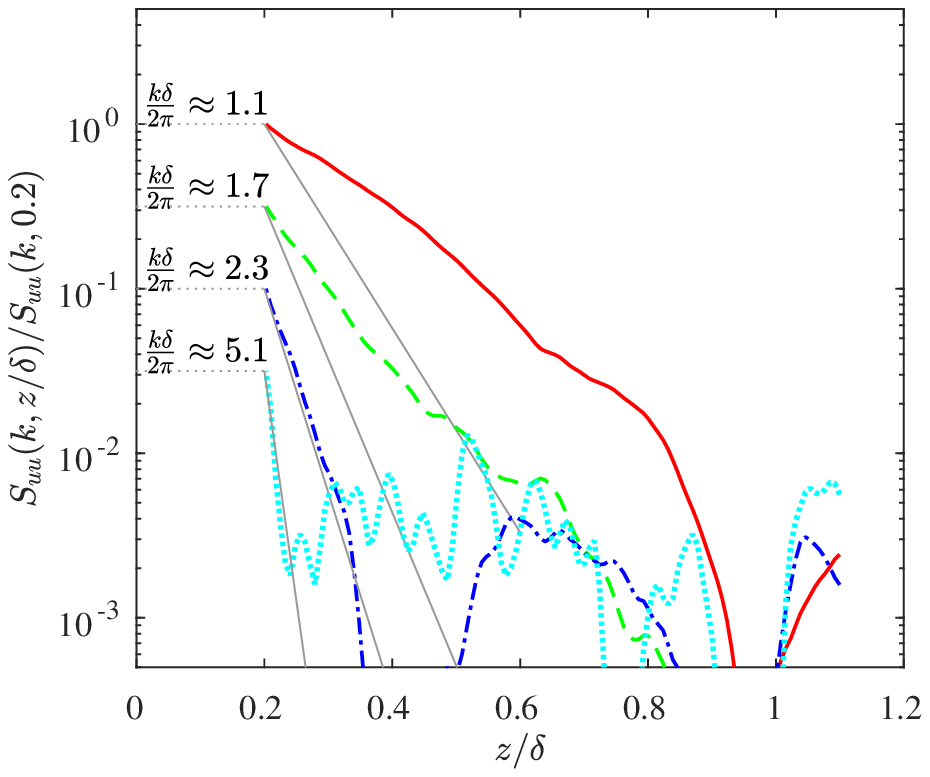}
		\includegraphics[width=0.42\textwidth]{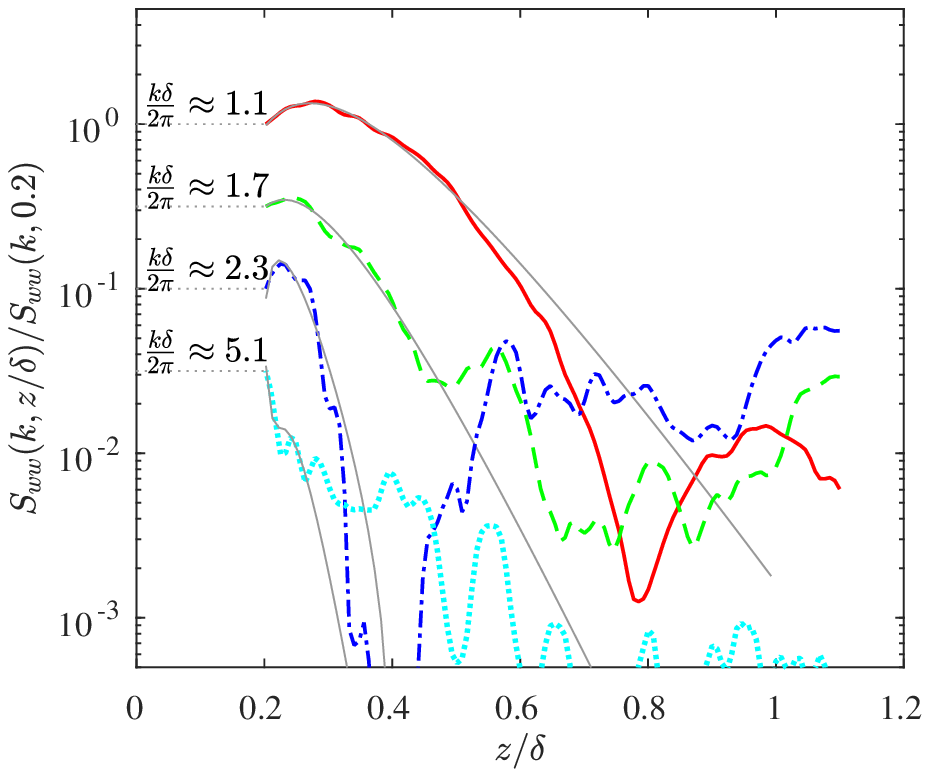}{\small (e)}
	\end{center}
	\caption{(continued from previous page) }\label{f:modedecay}
\end{figure}

Looking at the right panels of figure~\ref{f:modedecay}, it is observed that the fits of $S_{ww}$ match the data very well between $z/\delta =0.2$ and $z/\delta =0.6$ for all cases and modes. At higher wall distances, the measurements are saturated with noise, such that no meaningful comparison can be performed. Considering the matching of $\exp(-2 k_2 z)$ to $S_{uu}$ in the left panels of figure~\ref{f:modedecay}, the picture is more diverse. For the first two cases (panel a and b), a good agreement is found between $z/\delta =0.2$ and $z/\delta =0.3$ to 0.4. Beyond that, the noise level starts to dominate the measurements. For the other cases (panel c--d), an adequate matching for higher modes is generally found, i.e. in particular for $k\delta/(2\pi) = \delta/\ell_k >2$. For lower modes, the experimental slopes do not correspond well with the analytical slope, and are typically lower in absolute value.

\section{Discussion}\label{s:discussion}
Given the strong assumptions made in the development of the analytical solutions in \S\ref{s:analytical}, the correspondence with experiments is quite satisfactory. Differences between theory and data observed in figure~\ref{f:modedecay} for modes with $k\delta/(2\pi) = \delta/\ell_k < 2$ can be attributed to various factors. First of all, nonlinear effects can play a role, in particular since these modes tend to occur mainly for $S/\delta=0.88$, $1.2$, and $1.8$, which have much stronger dispersive stresses relative to the friction velocity (see figure~\ref{f:overview_stresses}) and refer to the developed derivation. Also, $\Gamma \approx 0$ and $U\approx U_\infty$ are strong simplifications: e.g., around $z/\delta = 0.2$, the velocity deficit is already about 30\% of the free-stream velocity.

A further assumption in the derivation of (\ref{eq:finaleqomega},\ref{eq:finalequ3}) is that the streamwise evolution of mean velocity components can be neglected in (\ref{eq:planeavNS}) and (\ref{eq:NSdispersive}). For $k_1\rightarrow 0$, this assumption does not hold. Consequently, the limit $k_1/k\rightarrow 0$ in (\ref{eq:expressionK}) is not viable, and only $\ell_k/\delta \rightarrow 0$ (and thus, $k\delta \rightarrow \infty$) may be expected to lead to (\ref{eq:solutionk10omw}). The picture in figure~\ref{f:modedecay} above is consistent with that, showing a good match between theory and results for $\delta/\ell_k > 2$. This is also in agreement with results from \cite{Vanderwel2015} and \cite{Hwang2018} that show that the excitation of secondary motions is most effective around a spanwise roughness spacing of  $S/\delta = \mathcal{O}(1)$, but decreases drastically when $S$ is decreased. 

The current analytical model may shed some further light on the characterization of roughness length scales in rough boundary layers. Whereas the classical roughness `height' is directly associated with the skin friction drag, the horizontal roughness length $\ell_k$ dictates how fast dispersive flow perturbations decay in the outer layer, supporting the notion of spanwise homogeneous outer flow behavior that is independent of the near-wall flow when $\ell_k$ is sufficiently small. Current findings may also be of interest for the development of rough-wall stress conditions for simulations, in particular in situations with multi-scale roughness (e.g. prevalent in the atmospheric boundary layer) in which only part of the roughness is resolved by the mesh \citep{Anderson2011}. 


Finally, the analytical solutions were obtained for $\Gamma=0$, and $\nu_t$ constant. Other, more involved solutions can be obtained by either solving (\ref{eq:finaleqomega},\ref{eq:finalequ3}) using more involved parametrizations of $\nu_e(z)$ in the outer layer of a boundary layer, or by directly solving the coupled system (\ref{eq:coupledeq1},\ref{eq:coupledeq2}) using an additional parametrization of $\Gamma(z)$ and $U(z)$. In this case, analytically tractable solutions may not anymore exist, but numerical solutions should be obtained without much complication. In this context, more extensive comparison with data, in particular including also roughness elements with $k_1\neq 0$ is also relevant. These are interesting topics for future research.


\appendix
\section{Elaboration of the linearized equations in Fourier space}

\subsection{Linearized equations in Fourier space}
Expressing the linearized equations (\ref{eq:linearizedcontinuity},\ref{eq:linearizedmomentum}) mode-by-mode using  (\ref{eq:Fourierspace1},\ref{eq:Fourierspace2}), etc.  (or formally using a Galerkin projection) leads to
\begin{align}
\imath k_1 \ft{u}_1 +\imath k_2 \ft{u}_2 + \Od[\ft{u}_3]{z}  = 0 \\
\imath k_1  U \ft{u}_1 + \ft{u}_3 \Gamma = - \imath k_1  \frac{\ft{p}}{\rho} + \tilde{F}_1 \\
\imath k_1  U \ft{u}_2  = - \imath k_2 \frac{\ft{p}}{\rho} + \tilde{F}_2 \\
\imath k_1  U \ft{u}_3  = -\frac{1}{\rho}\Pd[\ft{p}]{z} + \tilde{F}_3
\end{align}
Using (\ref{eq:expressu1},\ref{eq:expressu2}) to eliminate the continuity equation then leads to
\begin{align}
U \left(-\frac{k_1^2}{k^2} \Od[\ft{u}_3]{z} - \frac{k_1 k_2}{k^2} \ft{\omega}_3\right) + \ft{u}_3 \Gamma =& - \imath k_1  \frac{\ft{p}}{\rho} +   \tilde{F}_1 \label{eq:mom1appendix}\\
U (-\frac{k_1 k_2}{k^2} \Od[\ft{u}_3]{z} + \frac{k_1^2}{k^2} \ft{\omega}_3) =& - \imath k_2 \frac{\ft{p}}{\rho}+ \tilde{F}_2 \label{eq:mom2appendix} \\
\imath k_1  U \ft{u}_3  =& -\frac{1}{\rho}\Pd[\ft{p}]{z} + \tilde{F}_3 \label{eq:mom3appendix}
\end{align}
Using the first two equations, the pressure can be elaborated as 
\begin{align}
k^2 \frac{\ft{p}}{\rho} =&  \imath k_1 U \left(-\frac{k_1^2}{k^2} \Od[\ft{u}_3]{z} - \frac{k_1 k_2}{k^2} \ft{\omega}_3\right) + \imath k_1 \ft{u}_3 \Gamma \nonumber \\
& + \imath k_2  U (-\frac{k_1 k_2}{k^2} \Od[\ft{u}_3]{z} + \frac{k_1^2}{k^2} \ft{\omega}_3) - \imath k_1 \tilde{F}_1 - \imath k_2 \tilde{F}_2 \nonumber \\
=& \imath k_1  \ft{u}_3 \Gamma - \imath k_1 \tilde{F}_1 - \imath k_2 \tilde{F}_2 
\end{align}
Inserting back into (\ref{eq:mom1appendix}--\ref{eq:mom3appendix}) to eliminate the pressure, yields (\ref{eq:coupledeq1},\ref{eq:coupledeq2}) and one more equation that is linearly dependent.

\subsection{Elaboration of Reynolds forces and equations for $\Gamma=0$}

Elaboration of $\ft{G}_i$ using (\ref{eq:expressu1},\ref{eq:expressu2}) leads to 
\begin{align}
\ft{G}_1 =& \nu_t \left(\Od[^2\ft{u}_1]{z^2} - k^2  \ft{u}_1 \right) + \Od[\nu_t]{z} \left(\Od[\ft{u}_1]{z}+\imath k_1 \ft{u}_3\right) = \Od{z}\nu_t\Od[\ft{u}_1]{z} - \nu_t k^2 \ft{u}_1 + \imath \Od[\nu_t]{z}  k_1 \ft{u}_3 \nonumber \\
=& \imath  \Od{z}\left( \nu_t \left(\frac{k_1}{k^2} \Od[^2\ft{u}_3]{z^2} + \frac{k_2}{k^2} \Od[\ft{\omega}_3]{z} \right) \right) - \imath \nu_t k_1 \Od[\ft{u}_3]{z} - \imath \nu_t k_2 \ft{\omega}_3 + \imath \Od[\nu_t]{z}  k_1 \ft{u}_3 \\
\ft{G}_2 =& \imath  \Od{z}\left( \nu_t \left(\frac{k_2}{k^2} \Od[^2\ft{u}_3]{z^2} - \frac{k_1}{k^2} \Od[\ft{\omega}_3]{z} \right) \right) - \imath \nu_t k_2 \Od[\ft{u}_3]{z} + \imath \nu_t k_1 \ft{\omega}_3 + \imath \Od[\nu_t]{z}  k_2 \ft{u}_3 \\
\ft{G}_3 =& \Od{z}\nu_t\Od[\ft{u}_3]{z} - \nu_t k^2 \ft{u}_3 + \Od[\nu_t]{z} \Od[\ft{u}_3]{z}\\
\end{align}

Furthermore
\begin{align}
&k_2 \tilde{G}_1 - k_1 \tilde{G}_2 = \imath \Od{z} \nu_t \Od[\ft{\omega}_3]{z} - \imath \nu_t k^2 \ft{\omega}_3 \\
&-k_1 \Od[\tilde{G}_1]{z} -  k_2 \Od[\tilde{G}_2]{z} + \imath k^2 \tilde{G}_3 = -\imath \Od[^2]{z^2} \nu_t \Od[^2\ft{u}_3]{z^2} + 2\imath k^2 \Od[]{z} \nu_t \Od[\ft{u}_3]{z} - \imath k^2 \ft{u}_3  \left( \Od[^2 \nu_t]{z^2} + k^2\right) 
\end{align}
Inserting in (\ref{eq:coupledeq1}) and (\ref{eq:coupledeq2}), and further algebraic manipulation then leads to (\ref{eq:finaleqomega})~and~(\ref{eq:finalequ3}).

\end{document}